\documentclass[aps,twocolumn]{revtex4-1}

\usepackage{graphicx}

\usepackage{amssymb,amsfonts,amsmath}
\usepackage{multirow}
\usepackage{hyperref}

\begin{document}

\title{Unraveling the origin of exponential law in intra-urban human mobility}

\author{Xiao Liang}
\email{liangxiao@nlsde.buaa.edu.cn}
\author{Jichang Zhao}
\email{zhaojichang@nlsde.buaa.edu.cn}
\author{Li Dong}
\email{donglixp@gmail.com}
\author{Ke Xu}
\email[Corresponding author: ]{kexu@nlsde.buaa.edu.cn}
\affiliation{State Key Lab of Software Development Environment, Beihang University, Beijing 100191, P.R. China}


\begin{abstract} 
The vast majority of travel takes place within cities. Recently, new data has become available which allows for the discovery of urban mobility patterns which differ from established results about long distance travel. Specifically, the latest evidence increasingly points to exponential trip length distributions, contrary to the scaling laws observed on larger scales. In this paper, in order to explore the origin of the exponential law, we propose a new model which can predict individual flows in urban areas better. Based on the model, we explain the exponential law of intra-urban mobility as a result of the exponential decrease in average population density in urban areas. Indeed, both empirical and analytical results indicate that the trip length and the population density share the same exponential decaying rate.
\end{abstract}


\maketitle

Understanding human movement patterns is considered as a long-term fundamental but challenging task for decades. It is an essential component in urban planning \cite{Rozenfeld2008b, Jing2012, Zheng2011}, epidemics spreading \cite{Balcan22122009,Colizza2007,Wang2012} and traffic engineering \cite{Viboud21042006,Jung2012,Goh2012}. During the past few years, various mobile devices  (e.g., cellphones and GPS navigators) that support geolocation have been widely used in our daily life. As proxies, these devices record massive amounts of individual tracks, which provide a great opportunity to the research of human mobility.

In recent studies of human movements in large scale of space, including trips between counties or cities, it is found that human mobility patterns exhibit the L\'{e}vy walk characteristic, which corresponds to scale invariant step-lengths and is also observed in animals \cite{Viswanathan1996,Sims2008,Humphries2010,DeJager2011}. For instance, Brockmann et al. \cite{Brockmann2006} discover human travel displacements can be described by a power-law distribution by investigating the dispersal of bank notes in the United States. Gonz\'{a}lez et al. \cite{Gonzalez2008} study mobility patterns of mobile phone users in European countries and find that their travel distances are distributed according to a truncated power law. Moreover, the similar scaling laws are also observed in \cite{Song2010a} and \cite{Jiang2009} separately. Therefore, in order to understand the origin of  the scaling law, some researchers try to propose possible explanations from the viewpoint of individual movements  \cite{Han2011,Hu2011a,Jia2012}. 

However, regarding to the human movements in the urban area, many studies find that the human travel behavior could not be characterized by the scaling law but the exponential. For example, trajectories of passengers by taxis are investigated independently in three cities: Lisbon \cite{Veloso2011}, Beijing \cite{Liang2012} and Shanghai \cite{Peng2012}. And the three studies all suggest that trip distances obey exponential distributions rather than power-law ones. Bazzani et al. \cite{Bazzani2010} analyze daily round-trip lengths of private cars' drivers in Florence and reveal an exponential law of lengths, too. In addition, the distances of individual movement in the London subway are found to significantly deviate from the power-law distribution as well \cite{Roth2011}. Thus, Yan et al. \cite{Yan2012} think the exponential distribution is produced by a single means of transportation under Maxwell-Boltzmann statistics. But a more convincing evidence is that exponential distributions of intra-urban travel distances are demonstrated respectively in eight cities of Northeast China by analyzing the mobile phone data \cite{Kang2012}, which is not restricted to means of transportation. Moreover, with the aid of "checkins" of Foursquare users, Noulas et al. \cite{Noulas2011a} also discover that the trip-length distributions in different cities could not be approximated by power-law functions. They believe that urban human movements are driven by the distribution of places of interest (POIs) in the city. However, the conjecture might be challenged by the fact that the visit probabilities of POIs depend not only on their geographical locations, but also on their sizes and popularities.

Although more and more evidence demonstrates the exponential law in intra-urban human movements, the origin of this universal rule is still missed. Considering the significant role of the human mobility pattern in reality, it is essential to fill this vital gap. First, for most citizens, the majority of their trips occur in urban areas and just traverse small distances, while only few trips with large distances take place between counties or cities. Second, understanding the exponential law provides important guidance to model intra-urban movements. Compared to the scaling law, the exponential law implies lower probability of long travel distance, which could not be characterized by L\'{e}vy walks. Thus, modeling urban mobility according to the exponential law is more accordant with the real situations and is helpful for researches relevant to epidemic propagation control, wireless protocol designing and urban planning. Third, the discrepancy of trip-length distributions at different spatial scales can offer deeper insights into the consistency between individual and collective human mobility patterns. It is worth emphasizing that most empirical studies aforementioned are based on collective human trips. Yet there is no strong evidence that individual movements have the similar patterns with collective movements. For example, Yan et al. \cite{Yan2012} observe the absence of scaling law of travel distances at the individual level, though it indeed exists in aggregated trip lengths. Likewise, Noulas et al. \cite{Noulas2011a} discover that the power-law distribution of trip lengths is only found at the aggregated level. These recent discoveries all imply the multiformity of individual mobility patterns. Because of the diversity, it is worth further discussing whether it is reasonable that applying the properties observed in collective movements to model individual mobility patterns or exploring the source of the scaling law at the individual level. 

In this paper, we try to uncover the origin of the exponential law in intra-urban mobility from the perspective of collective movements. A new model is presented to predict individual flows between different regions with high fidelity, which could also reproduce the actual distributions of trip lengths in cities. Then from this model, we find that the traffic flux depends on the spatial population distribution heavily. Finally, both the empirical simulations and the analytical proof indicate that the exponential law is caused by the distribution of population density and they share the same decaying rate, too.

\section{Results}

\subsection{Modeling collective intra-urban mobility}
In this paper, human travel records were collected from four great cities by taxis, subways and surveys (see details in Methods \ref{mm:dataset}). It is demonstrated that the exponential law of collective human movements does exist in urban areas of cities (see Methods \ref{mm:distr}). 

In order to understand the exponential law of collective human mobility in urban areas, it is essential to model individual flows from one region to the other in a city. Although the gravity model \cite{Barthelemy2010} has already been applied widely to predict flows, including human travel \cite{Balcan22122009,Jung2012}, cargo ship movement \cite{Kaluza06072010} and telephone communications \cite{Krings2009}, it still has some flaws such as incompetence to explain the discrepancy of the numbers of individual flows in both directions between a pair of locations. Then in order to fix its disadvantages, Simini et al. \cite{Simini2012a} put forward the radiation model without parameters. In this model, the expected flux $\langle T_{ij}\rangle$ from location $i$ to $j$ is defined as $$\langle T_{ij}\rangle = T_i\frac{P_i P_j}{(P_i+P_{ij})(P_i+P_j+P_{ij})},$$ where $P_i$ and $P_j$ are the populations of location $i$ and $j$, $T_i$ is the number of trips starting from $i$ and $P_{ij}$ is the total population of locations (except $i$ and $j$) from which to $i$ the distances are less than or equal to $d_{ij}$ (the distance between $i$ and $j$). The model can predict population movements between counties or cities successfully \cite{Simini2012a}, but it is not clear whether the model applies to intra-urban movements as well.

Especially noteworthy is that in urban areas, it is difficult to obtain population distribution directly because of high mobility. Moreover, because people often move frequently for various purposes in cities, it is unsuitable to use resident population to model individual flows. Compared to the resident population, the average daily population occurring in a zone of city is more reasonable to characterize the urban mobility. Because it establishes a bond between human travel intensity and the function of the zone. So, in this paper, we regard the number of trips arriving at a zone as the population of the zone, which is proportional to the actual average daily population approximately.
 
\begin{figure}[htbp] 
  \includegraphics[scale=.2]{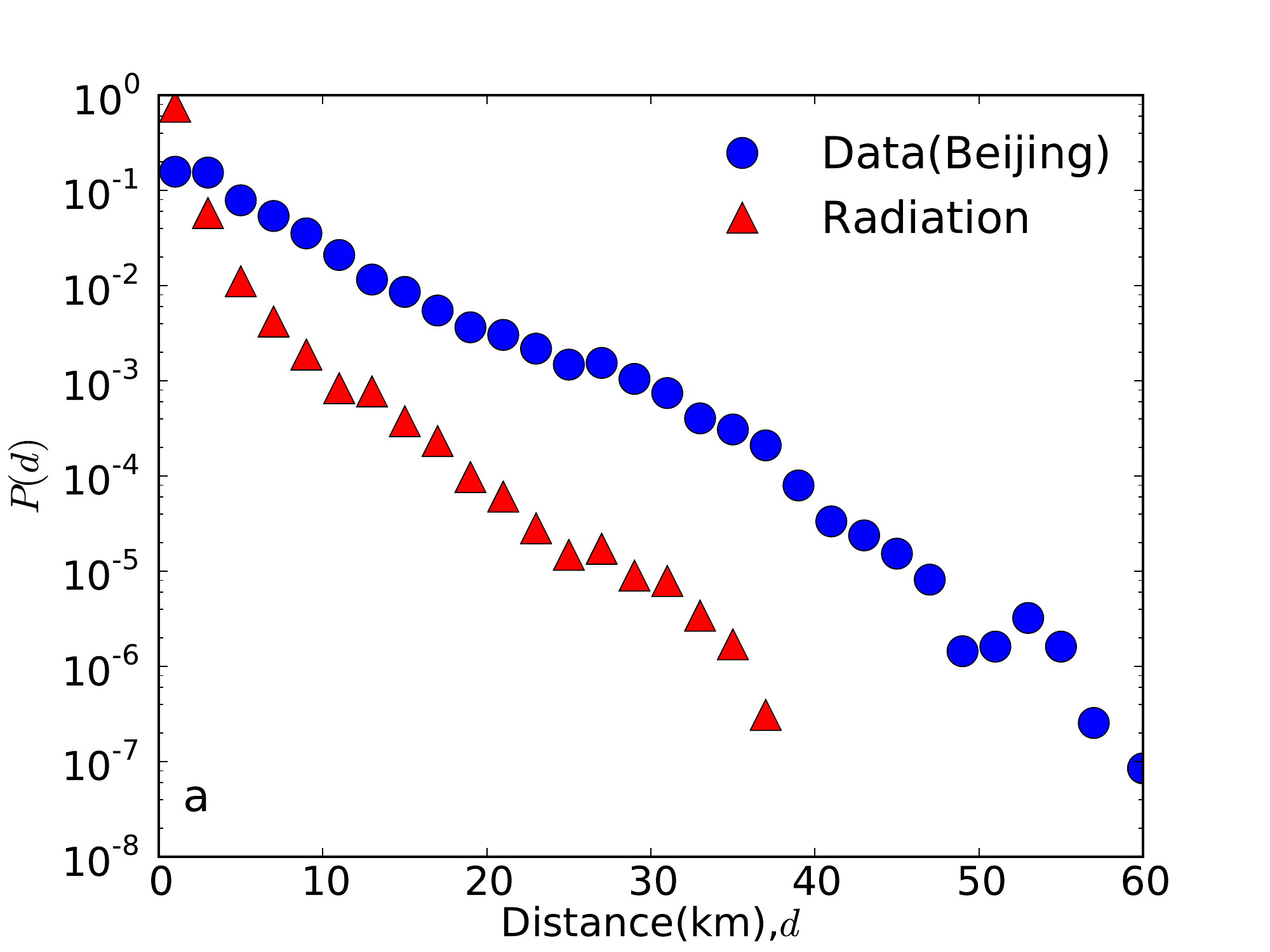}
  \includegraphics[scale=.2]{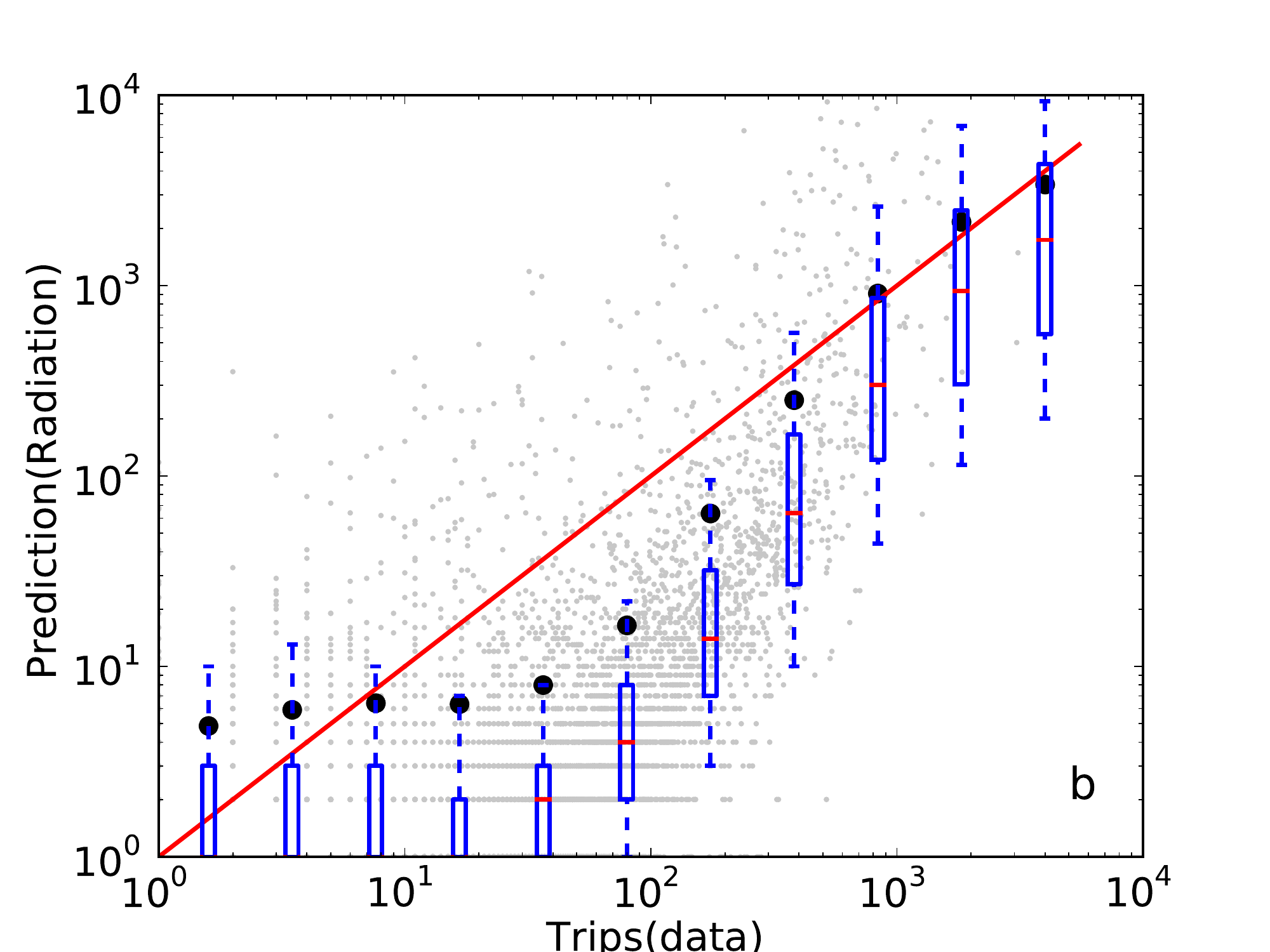}
  \caption{The simulations by the radiation model in Beijing. (a) The comparison of distance distributions between actual and simulated trips. (b) The prediction of traffic flows between regions. The grey points show the relationship between actual and predicted flux for ordered pairs of regions. The red line $y = x$ stands for the actual flows equal with predicted ones. The black points are the mean values of predicted flux in the bins. The ends of whisker represent the 9th and 91st percentile in the bins.}
  \label{fig:radiation}
\end{figure}

After calculating the population of zones, the results of simulation in Beijing by the radiation model is shown in Fig. \ref{fig:radiation}. From the figure, it seems that the predicted flux has a large deviation from the actual ones and the model underestimates the probabilities of trips with distances larger than 1 km. Similarly, the same phenomena can also be observed in other three cities, which are not included here. The possible reason to account for the incapability of radiation model is that there are different travel habits and preferences existing in trips at different scales of space. Therefore, it is necessary to consider a new model to understand intra-urban human mobility patterns.

Inspired by the gravity model \cite{Barthelemy2010}, we assume that the probability arriving at a location has a positive correlation with the population density of the location, but has a negative correlation with the Euclidean distance between the location and the originated location. Hence, in our model the probability of a trip reaching the location $n$, conditioned on starting from the location $m$, is defined as follows $$P(n|m) \propto \frac{\rho(n)}{f(d_{mn})},$$ where $\rho(\cdot)$ is the population density function and $f(d)$ is a function of distance between locations, which is usually given by two frequently used forms: power law and exponential. 

Likewise, as for regions, the probability of a trip arriving at the region $j$, conditioned on originating from the region $i$, is defined as  $$P(j|i) \propto \frac{P(j)}{f(d_{ij})},$$ where $P(\cdot)$ is the population of regions. Then the probability of a trip from region $i$ to $j$ can be derived as
\begin{eqnarray*}\label{formula:prob}
P(i\to j) &=& P_{norm}(i)P(j|i) \\
       &=&
       P_{norm}(i)\frac{P(j)/f(d_{ij})}{\sum_{k\neq i}{P(k)/f(d_{ik})}},
\end{eqnarray*} where $P_{norm}(\cdot)$ means the normalized population indicating the possibility of originating a trip from the region. Assuming $T$ is the total number of trips, the expected number of trips from region $i$ to $j$ can be concluded as
\begin{eqnarray*}\label{formula:pre}
\langle T_{ij}\rangle &=& T\cdot P(i\to j) \\
       &=&
       \frac{T}{\sum_{k\neq i}{P(k)/f(d_{ik})}}\frac{P_{norm}(i)P(j)}{f(d_{ij})} \\
& = & \frac{T}{M(i)}\frac{P_{norm}(i)P(j)}{f(d_{ij})},
\end{eqnarray*} where $M(i) = \sum_{k\ne i}{P(k)/f(d_{ik})}$.

As described in the gravity model, the number of trips from region $i$ to $j$ is equal with the one from region $j$ to $i$. However, that is not the case in our model because the values of $M(i)$ and $M(j)$ depend on geographic positions of $i$ and $j$ respectively, which are often not equal to each other. Therefore, it is more consistent with actual situations. 

\begin{figure}[htbp]
  \includegraphics[scale=.4]{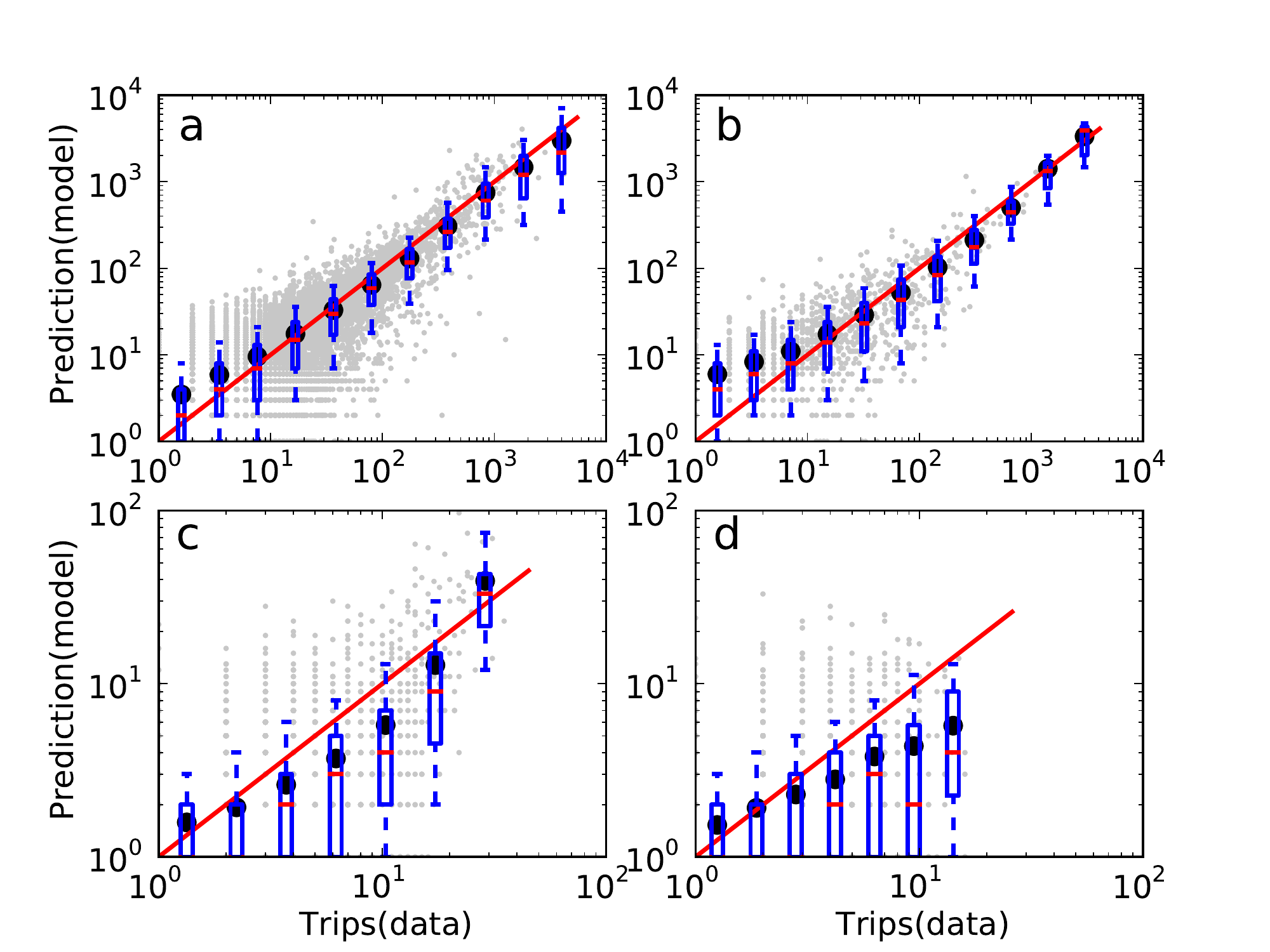}
  \caption{The relationships between actual and predicted traffic flux. (a)Beijing. (b)London. (c)Chicago. (d)Los Angeles.}
  \label{fig:prediction}
\end{figure}

\begin{figure}[htbp]
  \includegraphics[scale=.4]{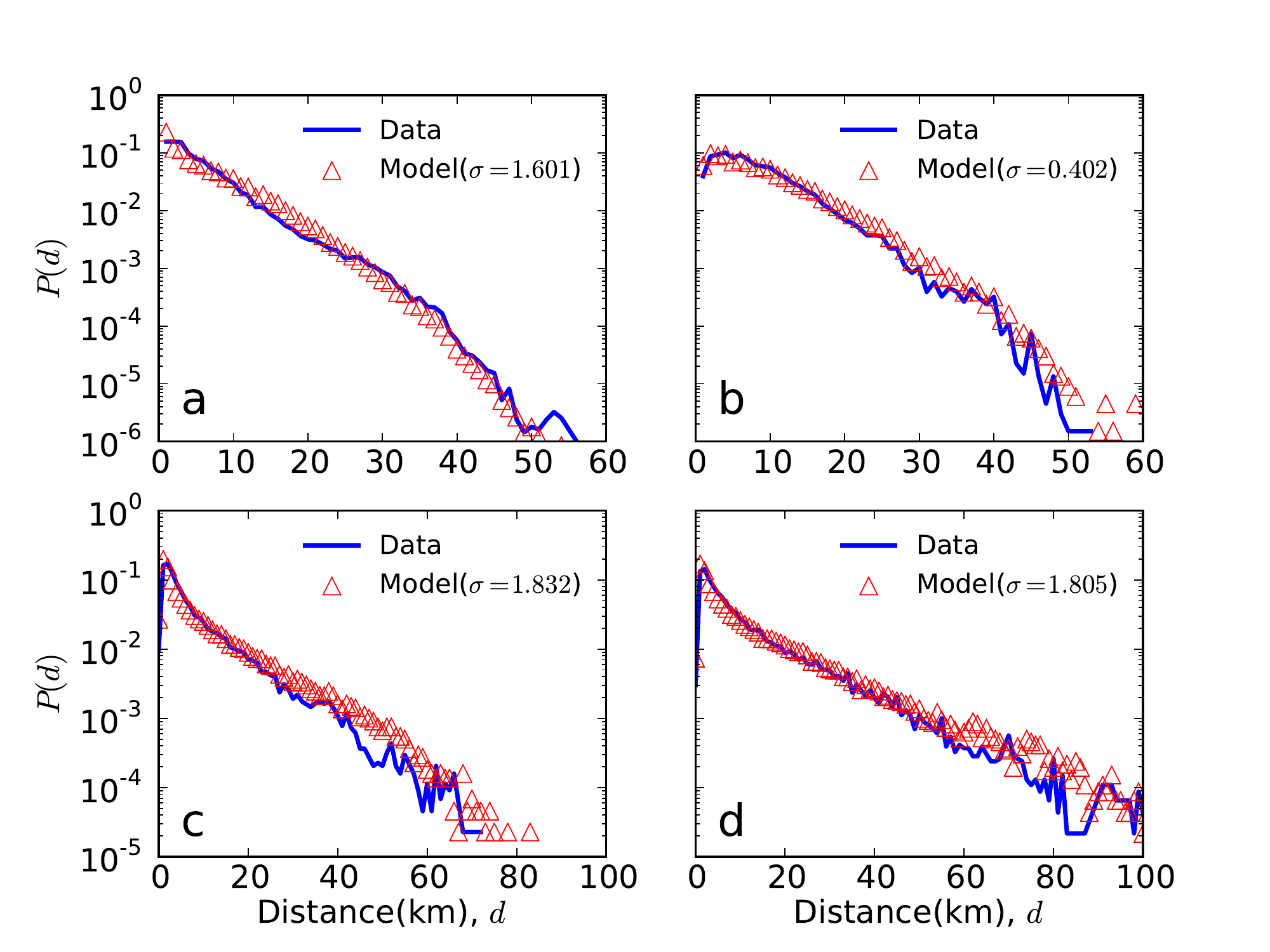}
  \caption{The distributions of actual and simulated trip distances in the four cities. The blue solid lines denote the actual traveling distance distributions. The red triangles represent the trip-length distributions simulated by our model. (a)Beijing. (b)London. (c)Chicago. (d)Los Angeles.}
  \label{fig:pred_distdistr}
\end{figure}

After given the actual trips, the parameter of  function $f(d)$ in our model can be determined by using the method of Maximum Likelihood Estimation (MLE). After inspecting the two forms of function $f(d)$ carefully, it is found that the power-law form  $f(d) = d^\sigma$ is much better. The power-law exponents $\sigma$ in our model for the four cities-Beijing, London, Chicago and Los Angeles-are 1.601, 0.402, 1.832 and 1.805 respectively. By using our model to simulate human travels in the four cities, the relationships between actual and predicted traffic flows are shown in Fig. \ref{fig:prediction}. From the subgraphs, it can be observed that the red lines $y = x$ almost lie between the 9th and the 91st percentiles in all bins except the last bin in Los Angeles, indicating that the model can predict the number of trips between regions accurately. Moreover, the comparisons of distributions of actual and simulated trip lengths are illustrated in Fig. \ref{fig:pred_distdistr}. It is discovered that the simulated distributions accord with the actual ones very well in the four cities. The fitted exponential parameters for the simulated distance distributions in the four cities are $0.1828\pm 0.0001$, $0.181\pm 0.0008$, $0.0903\pm 0.0017$ and $0.0696\pm 0.0012$ respectively, which are very close to the fitted values for actual trip-length distributions as shown in Table \ref{tb:mle_exp}. In summary, our model with the form of power law can be treated as an appropriate model to predict traffic flows in urban areas. 

\subsection{Analyzing the influence of population distribution}
\begin{figure}[htbp]
  \includegraphics[scale=.21]{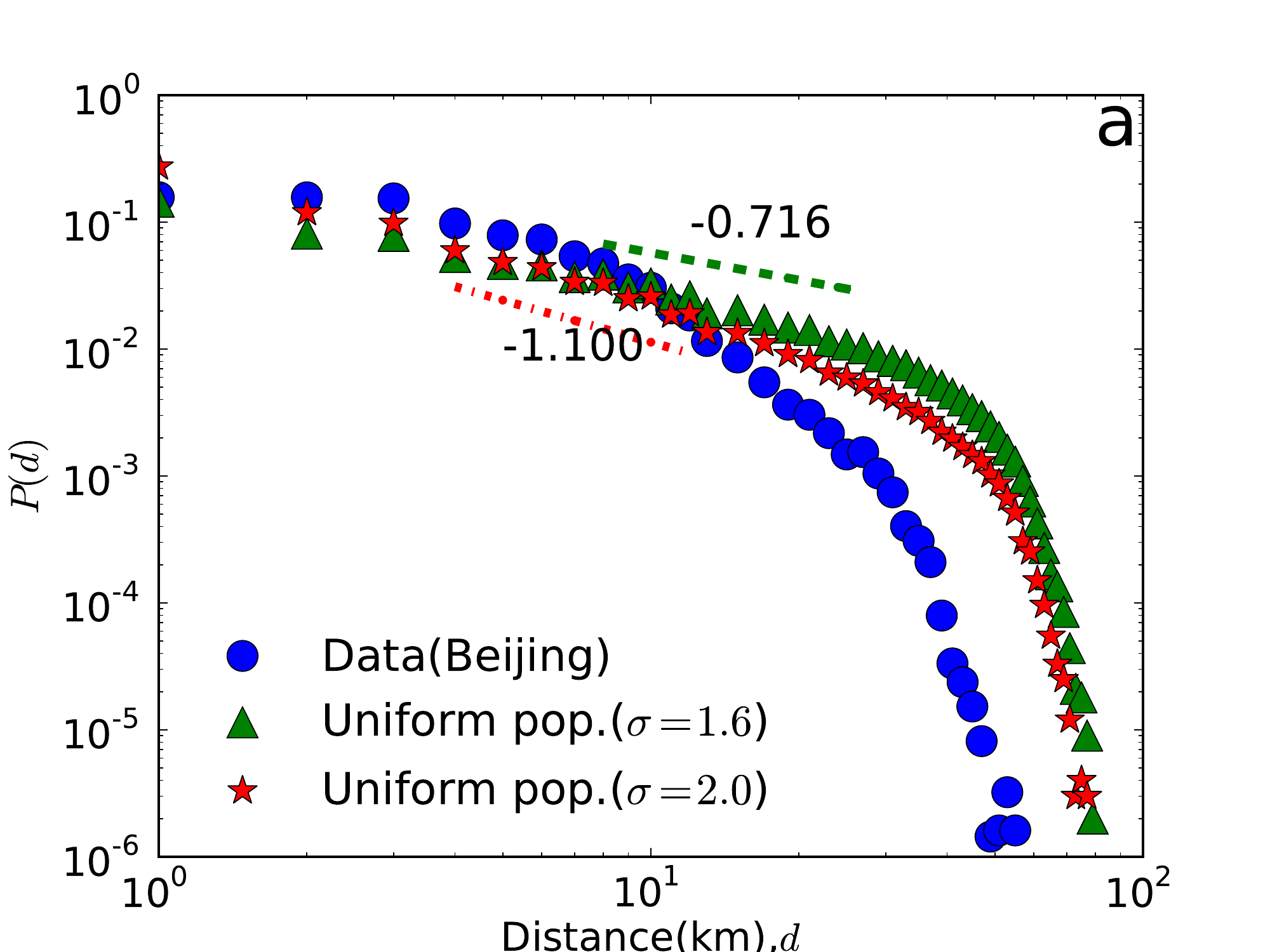}
  \includegraphics[scale=.21]{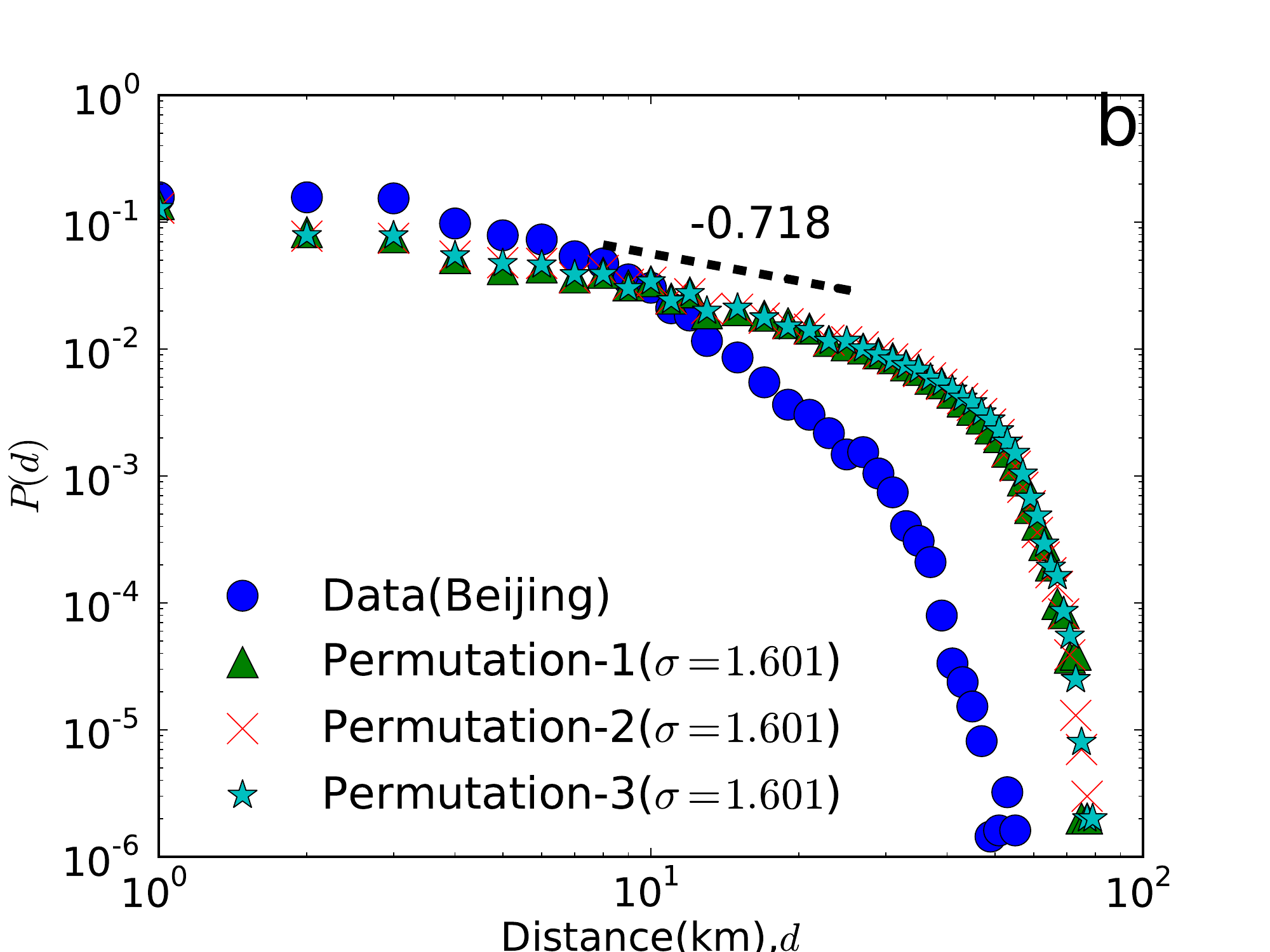}
  \caption{The simulated trip-length distributions with different spatial population distributions in Beijing. (a) Uniform population distributions with different model parameters. (b) Randomized permutation of  population numbers of cells.}
  \label{fig:pop_bj}
  \end{figure}

From the model, it is apparent that the spatial population distribution has an important impact on collective human mobility. Taking Beijing as an example, it is investigated that how the geographic population distribution could affect the trip-length distribution. First, considering the population distribution is uniform, individual trips could be predicted by our model with different parameters $\sigma$. As shown in Fig. \ref{fig:pop_bj}(a), contrary to the actual trip-length distribution, the simulated distance distributions accord to power laws with exponential cutoff very well and decay more slowly. The power-law exponents of the two simulated distributions are -0.716 for $\sigma = 1.6$ (green triangles) and -1.100 for $\sigma = 2.0$ (red stars), which approach to the analytical results $1-\sigma$ (see Methods \ref{mm:proof}). Second, remaining the distribution of population numbers of cells unchanged, three synthetic population distributions are generated by randomized rearranging the population numbers of cells. In simulations of human trips, the parameter $\sigma$ of our model is the fixed value 1.601, which is the same as the actual one in Beijing. In Fig. \ref{fig:pop_bj}(b), the simulated distributions are similar to each other and could be described by power-law distributions better. In summary, these demonstrate that not only the distribution of population numbers but also the layout of them could influence the trip-length distribution.

\begin{figure}[htbp]
  \includegraphics[scale=.4]{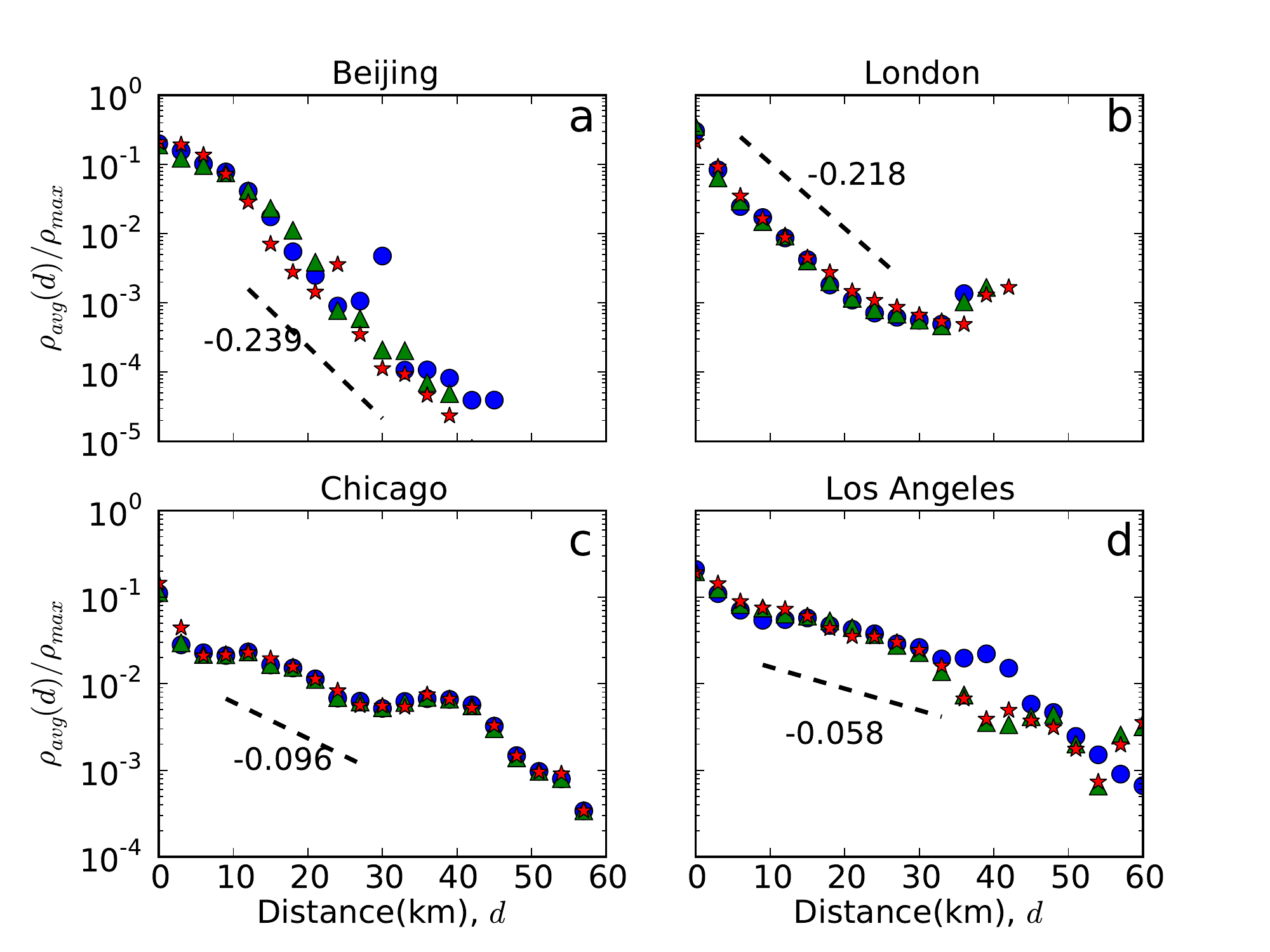}
  \caption{The normalized average density versus the distance to urban centers. For each city, three hot regions with high population densities are considered as urban centers. And the blue circles, green triangles and red stars denote the average densities, which are normalized by the maxima of densities, for selected urban centers respectively. The black dashed lines represent the decreasing rates of densities with distance.}
  \label{fig:pop_density}
  \end{figure}

Thus, it is necessary to study the spatial distribution of urban population. Here the relationships between the normalized average density and the distance to urban centers are plotted in Fig. \ref{fig:pop_density} for four cities (the details of calculating densities can be referred in Methods \ref{mm:pop}). From the graph, it can be observed that, for different selected centers of each city, the average urban densities have similar trends. More importantly, the densities for four cities all decay exponentially with the increase of distance to the urban centers. And it is worth noting that the declining slopes are not far from the exponents of exponential estimated from the corresponding distance distributions shown in Table \ref{tb:mle_exp}.

Assuming the density function $\rho(r)$ is a negative exponential function that depends on the distance $r$ to the center $$\rho(r)=Ce^{-\lambda r}(\lambda >0, 0 \leq r \leq R),$$ where C is a constant. The distance distribution $P(d)$ can be derived as $$C_1d^{1-\sigma}e^{-\lambda d} \leq P(d) \leq C_2d^{1-\sigma}e^{-\lambda d},$$ where $C_1$ and $C_2$ are constants (see the proof in Methods \ref{mm:proof}). Hence when $d > 1/\lambda$, the exponential section dominates and $P(d)$ begins to decay exponentially. 

\begin{figure}[htbp]
  \includegraphics[scale=.2]{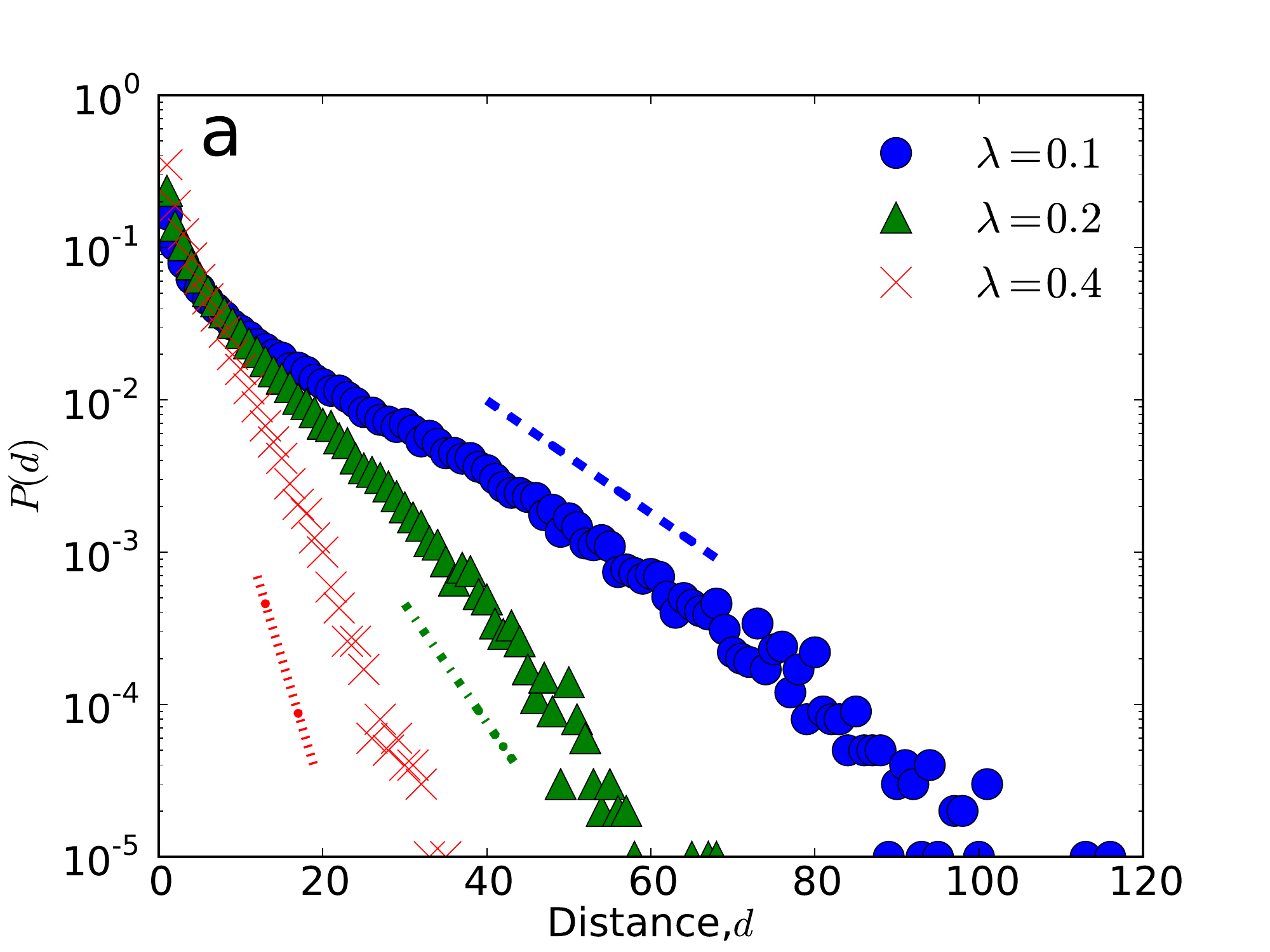}
  \includegraphics[scale=.2]{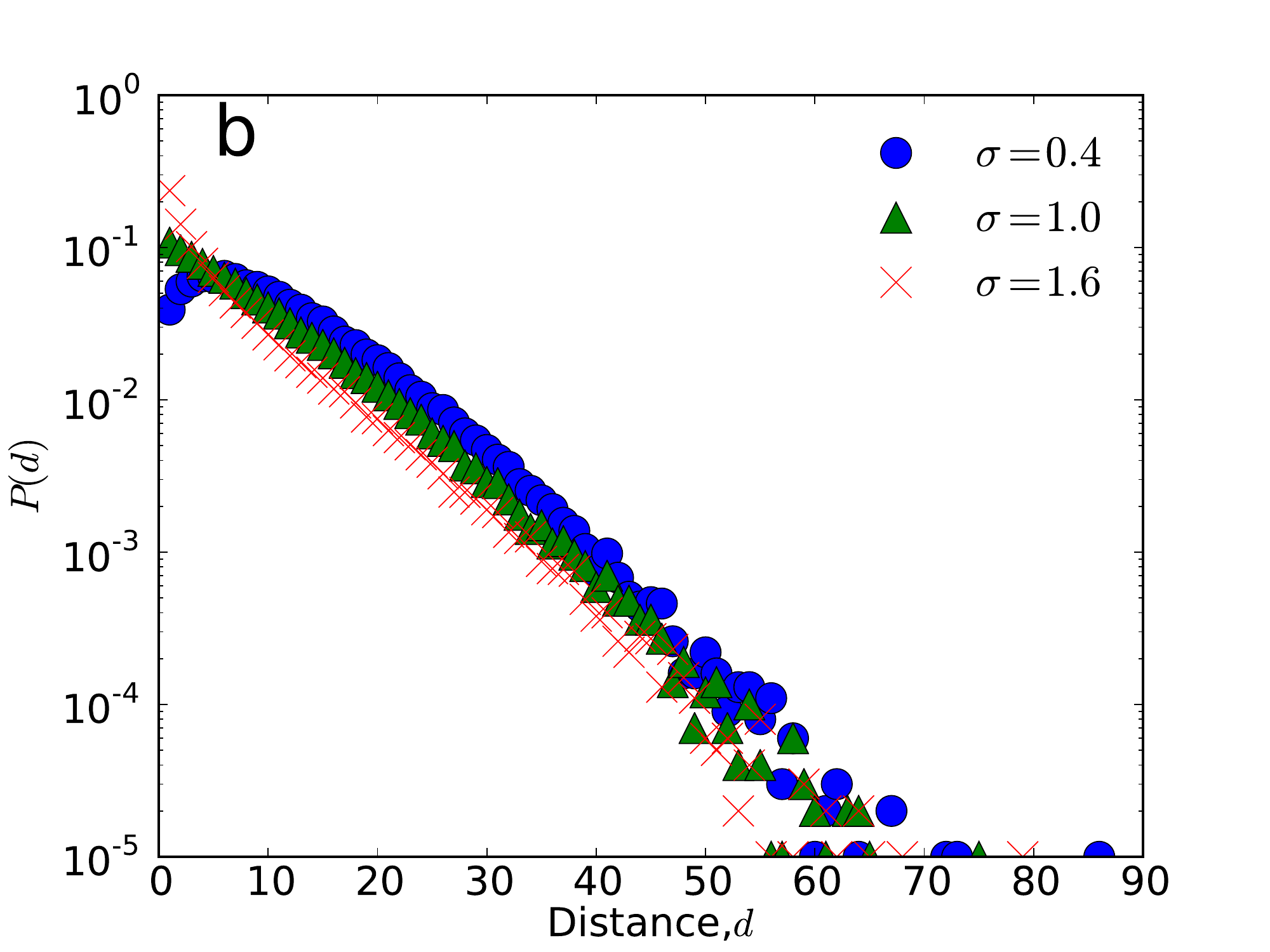}
  \caption{The simulations based on negative exponential distributions of population density. (a)Different population density distributions with the fixed model parameter $\sigma=1.6$. The simulated distance distributions, corresponding to different $\lambda$ (0.1, 0.2 and 0.4), decay exponentially with slope 0.085 (blue dashed line), 0.180 (green dash-dot line), 0.414 (red dotted line) respectively. (b) Different model parameters with the fixed population density distribution ($\lambda=0.2$).}
  \label{fig:sim_proof}
  \end{figure}

Then, it is aimed to verify the analytical result through simulating human trips based on our model further. And our model is simulated on grid cells whose size is $80\times 80$. When fixing the model parameter $\sigma$, as shown in Fig. \ref{fig:sim_proof}(a), the simulated trip-length distributions all exhibit exponential tails and the parameters of exponential distributions approach to the corresponding parameters $\lambda$ of population density distribution. From the Fig. \ref{fig:sim_proof}(b), the distance distributions have similar rates of exponential decay indicating that the model parameter $\sigma$ has little influence on the exponential tails of distributions when fixing the parameter $\lambda$ of the density function. In conclusion, the result of proof agrees with the simulations very well.

According to the analytical result, it could be explained why the exponential parameters of actual distance distributions are close to the ones of population density distributions in the four cities. Furthermore, in urban areas, the density usually decreases significantly leading to a large exponent $\lambda$, thus a short range of power-law section. Meanwhile, it must be noticed that the power of $d$ is often larger than -1 because of $\sigma < 2$ like the four cities in our datasets, which is such a slow power-law decay and obviously not a L\'{e}vy walk that has been observed in collective human movements in large scale of space. Therefore, it can explain the reason why the distance distribution of human trips in urban areas accords with an exponential distribution much better.

It is worth mentioning that the empirical density distributions in four cities is the same as the Clark's model \cite{Clark1951}, which is the most influential model for describing urban population density. Since then, some studies have proposed other mathematical forms for population density. For example, the inverse power function is employed by Smeed \cite{Smeed1963}. Though some controversies, Parr \cite{Parr1985} suggests that the negative exponential function is more appropriate to model the density in urban areas, while the inverse power function is more appropriate to model the variation of density in the urban fringe and hinterland. Therefore, the phase transition of population density function in different scales of space may be able to explain the different laws emerging in collective human mobility patterns.

\section{Discussion}

In the paper, it is aimed to understand the exponential law of intra-urban human mobility at the population level. The four travel datasets in urban areas of cities are analyzed, which further confirm the exponential law.  Through considering the travel flows between regions, it is clear that the radiation model is incapable to model collective human movements in urban areas. Because of this, a new model is proposed, which can predict traffic flows between regions very well. Based on our model, it is discovered that the average population density decreasing exponentially with distance to the urban center ultimately leads to the exponential law of collective human mobility patterns. Moreover, the difference of population distribution in different scales of space may be able to explain the different laws (power-law and exponential) in collective human movements. 

In fact, from the exponential law, it is hard to conclude that the trip lengths at the individual level follow a power-law distribution. It must be noted that most empirical studies about human mobility patterns are at the population level, but many models are aimed to explore the origin of scaling law at the individual level. \cite{Petrovskii24052011} demonstrated that a scale-free distribution of the aggregated movement lengths can also be obtained from individuals with different exponential distributions of movement lengths. A new evidence in human temporal dynamics \cite{Jiang2013} is that, although the aggregated intercall durations follow a power-law distribution, the durations at the individual level follow a power-law distribution for only a small number of individuals and a Weibull distribution for the majority. In addition, the research of human mobility is inspired by animals. There are yet some controversies \cite{Edwards2007,Edwards2012,Sims2012} on whether animals exhibit L\'{e}vy-like behaviour. Because of these, the individual mobility patterns should be considered carefully and more comprehensive human travel records are needed for deeply empirical analysis.

\section{Methods}
\subsection{Data descriptions}\label{mm:dataset}
\subsubsection{Beijing}
The dataset is about the taxis' GPS data generated by over 10 thousand taxis in Beijing, China, during three months ended on Dec. 31st, 2010 \cite{Liang2012}. Based on taxis' locations and statuses of occupation (with passengers or without passengers), trajectories of passengers can be observed. After dividing the urban areas of Beijing (inside the 6th Ring Road) into grid-like cells with  size $0.01^{\circ} \times 0.01^{\circ}$, a total of 11776743 trajectories between 3450 cells were extracted.

\subsubsection{London}
The dataset contains about 5\% samples of human trips by Tube in London, which were captured by Oyster cards during a week in November 2009 (Available online at \url{http://www.tfl.gov.uk/businessandpartners/syndication/}). In the dataset, the stations that a journey started or ended at were recorded. It was noticed that some stations were very close to each other, even less than 200 m. Because of too small area of regions, it could not reflect the regular mobility patterns between regions obviously. After merging some adjacent stations,  we obtained 183 voronoi cells based on stations and a total of 667584 trips between them.
\subsubsection{Chicago}
The dataset used here comes from the household travel tracker survey in Chicago Metropolitan areas conducted by Chicago Metropolitan Agency for Planning from January 2007 to February 2008 (Available online at \url{http://www.cmap.illinois.gov/travel-tracker-survey/}). The survey contained various kinds of information about households and travel activities of household members. Among them, the trips occurred in the Cook county were considered which is seen as the urban area of the city according to the population density. Then we extracted a total of 43881 trips between the 1314 zones which correspond to the census tracts.
\subsubsection{Los Angeles}
The Post-Census Regional Household Travel Survey, sponsored by the Southern California Association of Governments in 2001, was aimed to investigate human travel behavior in the Los Angeles region of California (Available online at \url{http://www.scag.ca.gov/travelsurvey/}). The region consisted of six counties. In terms of the survey data, it was paid more attention to the movements in the Los Angeles county. As a result, based on the census tracts, the county was divided into 2017 zones and a total of 46000 tracks were identified between these zones.

\subsection{Distance distributions in urban areas}\label{mm:distr}
There are three candidate models (power law, exponential and truncated Pareto) to be compared for describing the empirical distribution of travel distances. The parameters of these models are determined by the method of maximum likelihood estimation (MLE).  The Akaike weights are calculated, which are considered as relative likelihoods being the best model. The model with the largest Akaike weight should be selected as the best one. The details of methods can be found in \cite{Edwards2007, Edwards2012}.

\begin{figure}[htbp]
\includegraphics[scale=.4]{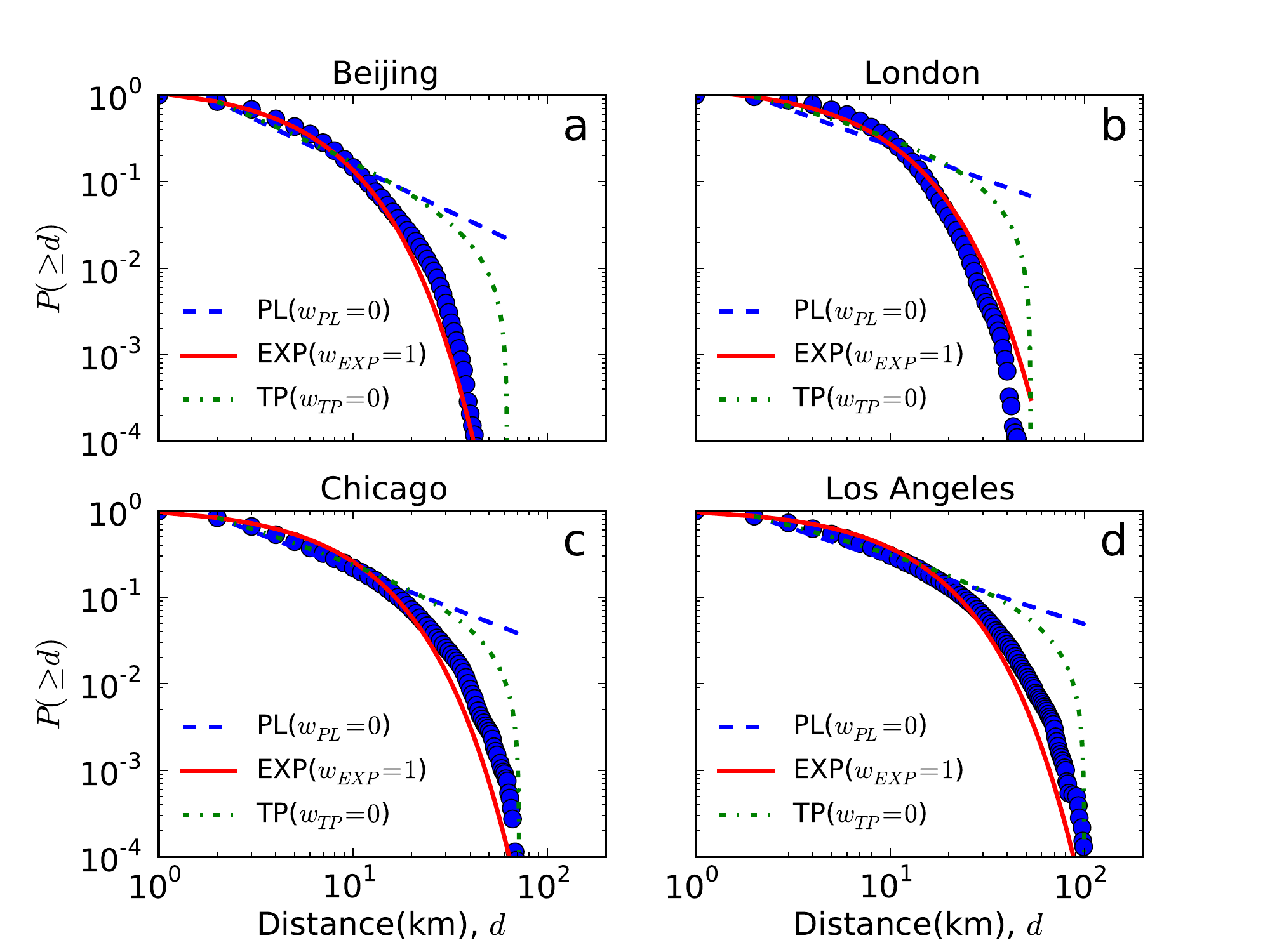}
\caption{The complementary cumulative distributions of travel distances in four cities. For each city, the fitted results of three models (power law (PL), exponential (EXP) and truncated Pareto (TP)) are plotted by the blue dashed line, the red solid line and the green dash-dot line respectively. $w_{PL},w_{EXP},w_{TP}$ are the Akaike weights for the three models.}
\label{fig:dist_distr}
\end{figure}

As shown in Fig. \ref{fig:dist_distr}, the distribution of trip lengths in each city is better fitted by the exponential rather than the power law or truncated Paretos. Moreover, the Akaike weight of exponential model ($w_{EXP}$ equals about 1.0) is significantly larger than the ones of the other two models in each city. These all demonstrate that the exponential law of collective human movements exists in urban areas of cities extensively. 

Furthermore, the fitted results of the exponential model in four cities are shown in Table \ref{tb:mle_exp}. The 95\% CI represents the 95\% confidence interval of the exponential parameter. And the goodness of fit (GOF) is measured by the Kolmogorov-Smirnov statistic. The smaller the value is, the more similar the empirical distance distribution and the fitted exponential model are.

\begin{table}[htbp]
\caption{The results of best fitted exponential model in four cities.}
\label{tb:mle_exp}
\begin{tabular}{cccc}
\hline
City & Parameter(EXP) & 95\% CI & GOF \\
\hline
Beijing & 0.2283 & (0.2282, 0.2284) & 0.0843 \\
London & 0.2061 & (0.2050, 0.2072) & 0.0122 \\
Chicago & 0.1084 & (0.1061, 0.1106) & 0.0133 \\
Los Angeles & 0.0774 & (0.0759, 0.0789) & 0.0102 \\
\hline
\end{tabular}
\end{table}

\subsection{Intra-urban population distribution}\label{mm:pop}
The spatial distribution of population in a city is often characterized by average population density with the distance to the urban center. As for the dataset of Beijing, the trips are in very fine granularity and there are similar-sized regions (cells) with small area. After selecting the cells with high densities as urban centers, the average densities with distance to centers are calculated easily. But, for the datasets of other three cities, there are coarse granularity of travels and irregular zones. It is not suitable to compute the average density directly. Therefore, assuming that population density in each zone is uniform, the urban area is divided into grid-like cells with size $0.005^{\circ}\times 0.005^{\circ}$. The population density of each cell is regard as the density of the zone in which the cell lies. Finally, the average densities with distance can be calculated based on these divided grid cells.

\subsection{Proof of the trip-length distribution}\label{mm:proof}
\begin{figure}[htbp]
\includegraphics[trim=120 300 10 300, scale=.7]{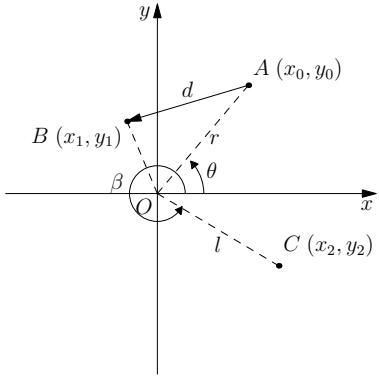}
\caption{Illustration to the proof of distance distribution.}
\label{fig:monocenter}
\end{figure}

As shown in Fig. \ref{fig:monocenter}, the center is the point $O$ and the non-increasing density distribution is $\rho(r)~(0\leq r\leq R)$, where $r$ is the distance to the center $O$ and $R$ is the size of a city. By using our model, we can estimate the displacement distribution of human movements as follows:
$$P(d) = \sum_{d(A,B)=d}{P(A\rightarrow B)}$$

Supporting continuity of the density distribution, it can also be written by
$$P(d) = \int\!\!\!\int \rho(A)
\frac{\int_{s:d(A,B)=d}{\rho(B)d^{-\sigma}\mathrm{d}s}}{\int\!\!\!\int
_{C\neq B} \rho(C)d(C,A)^{-\sigma} \mathrm{d}S_2}
\mathrm{d}S_1$$

Let $$I_1 = \int_{s:d(A,B)=d}{\rho(B)d^{-\sigma}\mathrm{d}s}$$
and $$I_2 = \int\!\!\!\int
_{C\neq B} \rho(C)d(C,A)^{-\sigma} \mathrm{d}S_2.$$

Therefore, 
\begin{eqnarray*}
I_1 &=& \int_{s:(x_1-x_0)^2 +(y_1-y_0)^2=d^2}{\rho(\sqrt{x_1^2+y_1^2})d^{-\sigma}\mathrm{d}s}\\
& & (x_1=x_0+d\cos{\alpha},y_1=y_0+d\sin{\alpha})\\
&=& \int_0^{2\pi} {d^{1-\sigma}\rho(\sqrt{x_0^2 + y_0^2 + d^2 + 2d(x_0\cos{\alpha} + y_0\sin{\alpha})})\mathrm{d}\alpha},
\end{eqnarray*}
\begin{eqnarray*}
I_2 &=& \int\!\!\!\int_{(x_2,y_2)\neq(x_0,y_0)}\frac{\rho(\sqrt{x_2^2+y_2^2})}{(\sqrt{(x_2-x_0)^2+(y_2-y_0)^2})^\sigma}\mathrm{d}x_2\mathrm{d}y_2\\
& & (x_2=l\cos{\beta}, y_2=l\sin{\beta})\\
&=& \int_0^{2\pi} \mathrm{d}\beta \int_{0}^{R} \frac{l\rho(l)}{(\sqrt{l^2+x_0^2+y_0^2-2l(x_0\cos{\beta}+y_0\sin{\beta})})^{\sigma}}\mathrm{d}l.\\ 
\end{eqnarray*}

Consequently, $P(d)$ can be represented as
\begin{eqnarray*}
P(d) &=& \int\!\!\!\int \rho(\sqrt{x_0^2+y_0^2})\frac{I_1}{I_2}\mathrm{d}x_0\mathrm{d}y_0\\
& & (x_0=r\cos{\theta},y_0=r\sin{\theta})\\
&=& \int_0^{2\pi} \mathrm{d}\theta \int_0^R r\rho(r)\frac{U}{V}\mathrm{d}r,
\end{eqnarray*}
where
\begin{eqnarray*}
U &=& \int_0^{2\pi} d^{1-\sigma}\rho(\sqrt{r^2+d^2+2dr\cos{(\theta-\alpha)}})\mathrm{d}\alpha,\\
V &=& \int_0^{2\pi} \mathrm{d}\beta \int_{0}^{R} l\rho(l){(\sqrt{l^2+r^2-2rl\cos{(\beta-\theta)}})^{-\sigma}}\mathrm{d}l.
\end{eqnarray*}

It is noticed that the denominator $V$ has nothing to do with $d$. Assuming the non-increasing density function $$\rho(r) = Ce^{-\lambda r}(\lambda \geq 0),$$ the numerator $U$ satisfies
$$ U \geq  C\int_0^{2\pi} d^{1-\sigma} e^{-\lambda (r+d)}\mathrm{d}\alpha = 2\pi Cd^{1-\sigma}e^{-\lambda r}e^{-\lambda d}.$$

In a similar way, 
$$ U \leq C\int_0^{2\pi} d^{1-\sigma}e^{-\lambda (d-r)}\mathrm{d}\alpha = 2\pi Cd^{1-\sigma}e^{\lambda r}e^{-\lambda d}.$$

As a result, when $\lambda = 0$, that is a uniform density function, $$ C_1d^{1-\sigma} \leq P(d) \leq C_2d^{1-\sigma}.$$ And when $\lambda > 0$, that is a negative exponential density funciton, $$C_1d^{1-\sigma}e^{-\lambda d} \leq P(d) \leq C_2d^{1-\sigma}e^{-\lambda d}.$$

\end{document}